
\def\Kel{\hskip 2pt K\hskip 3pt}
\def\etal{{\it et al.}\hskip 1.5pt}
\def\cf{{\it cf.}\hskip 1.5pt}
%
\newbox\hdbox%
\newcount\hdrows%
\newcount\multispancount%
\newcount\ncase%
\newcount\ncols
\newcount\nrows%
\newcount\nspan%
\newcount\ntemp%
\newdimen\hdsize%
\newdimen\newhdsize%
\newdimen\parasize%
\newdimen\thicksize%
\newdimen\thinsize%
\newdimen\tablewidth%
\newif\ifcentertables%
\newif\ifendsize%
\newif\iffirstrow%
\newif\iftableinfo%
\newtoks\dbt%
\newtoks\hdtks%
\newtoks\savetks%
\newtoks\tableLETtokens%
\newtoks\tabletokens%
\newtoks\widthspec%
%
%
\immediate\write15{%
-----> TABLE MACROS LOADED%
}%
%
%
\tableinfotrue%
\catcode`\@=11
\def\tstrut{\vrule height16pt depth6pt width0pt}%
\def\|{|}
\def\tablerule{\noalign{\hrule height\thinsize depth0pt}}%
\thicksize=1.5pt
\thinsize=0.6pt
\def\thickrule{\noalign{\hrule height\thicksize	depth0pt}}%
\def\ctr#1{\hfil\ #1\hfil}%
%
%
%
\tablewidth=-\maxdimen%
\def\tabskipglue{0pt plus 1fil minus 1fil}%
%
%
\centertablestrue%
%
%
%
%
\parasize=4in%
\gdef\ARGS{########}
\gdef\headerARGS{####}
\def\@mpersand{&}
{\catcode`\|=13
\gdef\letbarzero{\let|0}
\gdef\letbartab{\def|{&&}}
}
{\def\ampskip{&\omit\hfil&}
\catcode`\&=13
\let&0
\xdef\letampskip{\def&{\ampskip}}
}
\def\begintable{
   \begingroup%
   \catcode`\|=13\letbartab%
   \catcode`\&=13\letampskip%
   \def\multispan##1{
      \omit \mscount##1%
      \multiply\mscount\tw@\advance\mscount\m@ne%
      \loop\ifnum\mscount>\@ne \sp@n\repeat%
   }
   \def\|{%
      &\omit\widevline&%
   }%
   \ruledtable
}
\long\def\ruledtable#1\endtable{%
%
%
%
   \offinterlineskip
   \tabskip 0pt
   \def\widevline{\vrule width\thicksize}
   \def\endrow{\@mpersand\omit\hfil\crnorm\@mpersand}%
   \def\crthick{\@mpersand\crnorm\thickrule\@mpersand}%
   \def\crnorule{\@mpersand\crnorm\@mpersand}%
   \let\nr=\crnorule
   \def\endtable{\@mpersand\crnorm\thickrule}%
   \let\crnorm=\cr
%
%
   \edef\cr{\@mpersand\crnorm\tablerule\@mpersand}%
   \the\tableLETtokens
%
%
   \tabletokens={&#1}
%
%
   \countROWS\tabletokens\into\nrows%
   \countCOLS\tabletokens\into\ncols%
%
%
   \advance\ncols by -1%
   \divide\ncols by 2%
   \advance\nrows by 1%
%
%
   \iftableinfo	%
      \immediate\write16{[Nrows=\the\nrows, Ncols=\the\ncols]}%
   \fi%
%
%
   \ifcentertables
      \line{
      \hss
   \else %
      \hbox{%
   \fi
      \vbox{%
	 \makePREAMBLE{\the\ncols}
	 \edef\next{\preamble}
	 \let\preamble=\next
	 \makeTABLE{\preamble}{\tabletokens}
      }
      \ifcentertables \hss}\else }\fi
   \endgroup
   \tablewidth=-\maxdimen
}
\def\makeTABLE#1#2{
   {
   \let\ifmath0
   \let\header0
   \let\multispan0
%
%
   \ifdim\tablewidth>-\maxdimen	%
 \widthspec=\expandafter{\expandafter t\expandafter o%
 \the\tablewidth}%
   \else %
      \widthspec={}%
   \fi %
   \xdef\next{
      \halign\the\widthspec{%
      #1
      \noalign{\hrule height\thicksize depth0pt}
      \the#2\endtable
%
      }
   }
   }
   \next
}
\def\makePREAMBLE#1{
   \ncols=#1
   \begingroup
   \let\ARGS=0
   \edef\xtp{\widevline\ARGS\tabskip\tabskipglue%
   &\tstrut\ctr{\ARGS}}
   \advance\ncols by -1
   \loop
      \ifnum\ncols>0 %
      \advance\ncols by	-1%
      \edef\xtp{\xtp&\vrule width\thinsize\ARGS&\ctr{\ARGS}}%
   \repeat
   \xdef\preamble{\xtp&\widevline\ARGS\tabskip0pt%
   \crnorm}
   \endgroup
}
\def\countROWS#1\into#2{
   \let\countREGISTER=#2%
   \countREGISTER=0%
   \expandafter\ROWcount\the#1\endcount%
}%
\def\ROWcount{%
   \afterassignment\subROWcount\let\next= %
}%
\def\subROWcount{%
   \ifx\next\endcount %
      \let\next=\relax%
   \else%
      \ncase=0%
      \ifx\next\cr %
	 \global\advance\countREGISTER by 1%
	 \ncase=0%
      \fi%
      \ifx\next\endrow %
	 \global\advance\countREGISTER by 1%
	 \ncase=0%
      \fi%
      \ifx\next\crthick	%
	 \global\advance\countREGISTER by 1%
	 \ncase=0%
      \fi%
      \ifx\next\crnorule %
	 \global\advance\countREGISTER by 1%
	 \ncase=0%
      \fi%
      \ifx\next\header %
	 \ncase=1%
      \fi%
      \relax%
      \ifcase\ncase %
	 \let\next\ROWcount%
      \or %
	 \let\next\argROWskip%
      \else %
      \fi%
   \fi%
   \next%
}
\def\counthdROWS#1\into#2{%
\dvr{10}%
   \let\countREGISTER=#2%
   \countREGISTER=0%
\dvr{11}%
\dvr{13}%
   \expandafter\hdROWcount\the#1\endcount%
\dvr{12}%
}%
\def\hdROWcount{%
   \afterassignment\subhdROWcount\let\next= %
}%
\def\subhdROWcount{%
   \ifx\next\endcount %
      \let\next=\relax%
   \else%
      \ncase=0%
      \ifx\next\cr %
	 \global\advance\countREGISTER by 1%
	 \ncase=0%
      \fi%
      \ifx\next\endrow %
	 \global\advance\countREGISTER by 1%
	 \ncase=0%
      \fi%
      \ifx\next\crthick	%
	 \global\advance\countREGISTER by 1%
	 \ncase=0%
      \fi%
      \ifx\next\crnorule %
	 \global\advance\countREGISTER by 1%
	 \ncase=0%
      \fi%
      \ifx\next\header %
	 \ncase=1%
      \fi%
\relax%
      \ifcase\ncase %
	 \let\next\hdROWcount%
      \or%
	 \let\next\arghdROWskip%
      \else %
      \fi%
   \fi%
   \next%
}%
{\catcode`\|=13\letbartab
\gdef\countCOLS#1\into#2{%
   \let\countREGISTER=#2%
   \global\countREGISTER=0%
   \global\multispancount=0%
   \global\firstrowtrue
   \expandafter\COLcount\the#1\endcount%
   \global\advance\countREGISTER by 3%
   \global\advance\countREGISTER by -\multispancount
}%
\gdef\COLcount{%
   \afterassignment\subCOLcount\let\next= %
}%
{\catcode`\&=13%
\gdef\subCOLcount{%
   \ifx\next\endcount %
      \let\next=\relax%
   \else%
      \ncase=0%
      \iffirstrow
	 \ifx\next& %
	    \global\advance\countREGISTER by 2%
	    \ncase=0%
	 \fi%
	 \ifx\next\span	%
	    \global\advance\countREGISTER by 1%
	    \ncase=0%
	 \fi%
	 \ifx\next| %
	    \global\advance\countREGISTER by 2%
	    \ncase=0%
	 \fi
	 \ifx\next\|
	    \global\advance\countREGISTER by 2%
	    \ncase=0%
	 \fi
	 \ifx\next\multispan
	    \ncase=1%
	    \global\advance\multispancount by 1%
	 \fi
	 \ifx\next\header
	    \ncase=2%
	 \fi
	 \ifx\next\cr	    \global\firstrowfalse \fi
	 \ifx\next\endrow   \global\firstrowfalse \fi
	 \ifx\next\crthick  \global\firstrowfalse \fi
	 \ifx\next\crnorule \global\firstrowfalse \fi
      \fi
\relax
      \ifcase\ncase %
	 \let\next\COLcount%
      \or %
	 \let\next\spancount%
      \or %
	 \let\next\argCOLskip%
      \else %
      \fi %
   \fi%
   \next%
}%
\gdef\argROWskip#1{%
   \let\next\ROWcount \next%
}
\gdef\arghdROWskip#1{%
   \let\next\ROWcount \next%
}
\gdef\argCOLskip#1{%
   \let\next\COLcount \next%
}
}
}
\def\spancount#1{
   \nspan=#1\multiply\nspan by 2\advance\nspan by -1%
   \global\advance \countREGISTER by \nspan
   \let\next\COLcount \next}%
\def\dvr#1{\relax}%
\def\header#1{%
\dvr{1}{\let\cr=\@mpersand%
\hdtks={#1}%
\counthdROWS\hdtks\into\hdrows%
\advance\hdrows	by 1%
\ifnum\hdrows=0	\hdrows=1 \fi%
\dvr{5}\makehdPREAMBLE{\the\hdrows}%
\dvr{6}\getHDdimen{#1}%
{\parindent=0pt\hsize=\hdsize{\let\ifmath0%
\xdef\next{\valign{\headerpreamble #1\crnorm}}}\dvr{7}\next\dvr{8}%
}%
}\dvr{2}}
\def\makehdPREAMBLE#1{
\dvr{3}%
\hdrows=#1
{
\let\headerARGS=0%
\let\cr=\crnorm%
\edef\xtp{\vfil\hfil\hbox{\headerARGS}\hfil\vfil}%
\advance\hdrows	by -1
\loop
\ifnum\hdrows>0%
\advance\hdrows	by -1%
\edef\xtp{\xtp&\vfil\hfil\hbox{\headerARGS}\hfil\vfil}%
\repeat%
\xdef\headerpreamble{\xtp\crcr}%
}
\dvr{4}}
\def\getHDdimen#1{%
\hdsize=0pt%
\getsize#1\cr\end\cr%
}
\def\getsize#1\cr{%
\endsizefalse\savetks={#1}%
\expandafter\lookend\the\savetks\cr%
\relax \ifendsize \let\next\relax \else%
\setbox\hdbox=\hbox{#1}\newhdsize=1.0\wd\hdbox%
\ifdim\newhdsize>\hdsize \hdsize=\newhdsize \fi%
\let\next\getsize \fi%
\next%
}%
\def\lookend{\afterassignment\sublookend\let\looknext= }%
\def\sublookend{\relax%
\ifx\looknext\cr %
\let\looknext\relax \else %
   \relax
   \ifx\looknext\end \global\endsizetrue \fi%
   \let\looknext=\lookend%
    \fi	\looknext%
}%
%
%
\def\tablelet#1{%
   \tableLETtokens=\expandafter{\the\tableLETtokens #1}%
}%
\catcode`\@=12
\def\refset{\parindent=0pt\hangafter=1\hangindent=1em}
\magnification=1200
\hsize=6.00truein
\hoffset=1.20truecm
\newcount\eqtno
\eqtno = 1
\parskip 3pt plus 1pt minus .5pt
\baselineskip 20pt plus .1pt
\centerline{ \  }
\vskip 0.35in
\centerline{\bf A HYDRODYNAMIC APPROACH TO COSMOLOGY:}
\centerline{\bf THE MIXED DARK MATTER COSMOLOGICAL SCENARIO}
\vskip 1.0in
\centerline{Renyue Cen and Jeremiah P. Ostriker}
\centerline{\it Princeton University Observatory}
\centerline{\it Princeton, NJ 08544 USA}
\vskip 2.0in
\centerline{Submitted to {\it The Astrophysical Journal}, Aug 27, 1993}
\vskip 0.3in
\centerline{Jan 25, 1993}
\vskip 0.7in
\vfill\eject

\centerline{\bf ABSTRACT}

We compute
the evolution of spatially flat, mixed
cold and hot dark matter (``MDM") models
containing both baryonic matter and two kinds of dark matter.
Hydrodynamics is treated with
a highly developed Eulerian
hydrodynamic code [see Cen (1992)].
A standard Particle-Mesh (PM) code is also used in parallel
to calculate the motion of the dark matter components.
We adopt the following parameters:
$h\equiv H_0/100{\rm km s}^{-1} {\rm Mpc}^{-1}=0.5$,
$\Omega_{cold} =0.64$,
$\Omega_{hot} =0.3$ and
$\Omega_{b} =0.06$
with amplitude of the perturbation spectrum
fixed by the COBE DMR measurements (Smoot \etal 1992) being $\sigma_8=0.67$.
Four different boxes are simulated with box sizes of
$L=(64, 16, 4, 1) h^{-1}$Mpc, respectively,
the two small boxes providing good resolution but little valid information
due to the absence of large-scale power.
We use $128^3\sim 10^{6.3}$ baryonic cells,
$128^3$ cold dark matter particles
and $2\times 128^3$ hot dark matter particles.
In addition to the dark matter we follow separately
six baryonic species (H, H$^+$, He, He$^+$, He$^{++}$, e$^{-}$)
with allowance for both (non-equilibrium)
collisional and radiative ionization in every cell.
The background radiation field is also followed in detail with
allowance made for both continuum and line processes,
to allow non-equilibrium heating and cooling
processes to be followed in detail.
The mean final Zeldovich-Sunyaev $y$ parameter
is estimated to be $\bar y=(5.4\pm 2.7)\times 10^{-7}$,
below currently attainable observations,
with a rms fluctuation of approximately
$\bar{\delta y}=(6.0\pm 3.0)\times 10^{-7}$ on arc minute scales.

The rate of galaxy formation peaks at an even later
epoch ($z\sim 0.3$) than in the standard ($\Omega=1$, $\sigma_8=0.67$)
CDM model ($z\sim 0.5$)
and, at a redshift of z=4
is nearly a factor of a hundred lower than for the CDM model
with the same value of $\sigma_8$.
With regard to mass function, the smallest
objects are stablized
against collapse by thermal energy:
the mass-weighted mass spectrum has a broad peak in the vicinity of
$m_b=10^{9.5}M_\odot$
with a reasonable fit to the Schecter luminosity
function if the baryon mass to blue light ratio is approximately $4$.

In addition, one very large PM simulation was made in a box
with size ($320h^{-1}$Mpc) containing $3\times 200^{3}=10^{7.4}$
particles. Utilizing this simulation we find that the model
yields a cluster mass function which is about
a factor of 4 higher than observed but a cluster-cluster correlation
length lower by a factor of 2 than what is observed but
both are closer to observations than in the COBE normalized CDM model.
The one dimensional pairwise velocity dispersion is
$605\pm 8$km/s at $1h^{-1}$ separation, lower than that
of the CDM model normalized to COBE, but
still significantly
higher than observations (Davis \& Peebles 1983).
A plausible velocity
bias $b_v=0.8\pm 0.1$ on this scale
will reduce but not remove the discrepancy.
The velocity auto-correlation function has a coherence
length of $40h^{-1}$Mpc, which is somewhat lower than
the observed counterpart.
In all these respects the model would be improved by decreasing
the cold fraction of the dark matter
and could be brought into agreement
with these constraints for a somewhat smaller value
of $\Omega_{CDM}/(\Omega_{CDM}+\Omega_{HDM})$.
But formation of galaxies and clusters of galaxies
is much later in this model than in COBE normalized CDM,
perhaps too late.
To improve on these constraints a larger ratio of
$\Omega_{CDM}/(\Omega_{CDM}+\Omega_{HDM})$ is required than the value
$0.67$ adopted here.
It does not seem possible to find a value for
this ratio which would satisfy all tests.

Overall, the model is similar
both on large and intermediate scales
to the standard CDM model
normalized to the same value of $\sigma_8$,
but the problem with regard to late formation of galaxies
is more severe in this model than in that CDM model.
Adding hot dark matter significantly improves
the ability of COBE normalized CDM scenario to
fit existing observations,
but the model is in fact not as good as the CDM model
with the same $\sigma_8$ and is still
probably unsatisfactory with regard to several critical tests.

\vskip 0.7cm
\noindent
Cosmology: large-scale structure of Universe
-- cosmology: theory
-- galaxies: clustering
-- galaxies: formation
-- hydrodynamics
\vfill\eject

\centerline{1. INTRODUCTION}

In a series of papers, we have used
a highly developed three dimensional hydrodynamic Eulerian code (Cen 1992)
to examine the evolution of baryonic matter as well as dark matter
in different model universes
(standard gaussian CDM and HDM models, Cen \& Ostriker 1992a(=CO92), 1992b;
texture-seeded CDM and HDM models, Cen \etal 1991;
tilted CDM model, Cen \& Ostriker 1993a;
PBI model, Cen, Ostriker \& Peebles 1993;
CDM$+\Lambda$ model, Cen, Gnedin \& Ostriker 1993).
All were treated with the same code and the same numerical resolution.
This paper is the last of this series.
We study here the hydrodynamic properties of the mixed
dark matter cosmological scenario which has been recently re-examined
(\cf Davis, Summers, \& Schlegel 1992=DSS herefter;
Taylor \& Rowan-Robinson 1992=TR hereafter;
Klypin \etal 1993=KHPR hereafter)
as a variant of the standard cold dark matter scenario.
The idea for such a model dates as far back as 1984
(\cf KHPR for a survey of the literature),
but recent observations of large-scale structure
have led to renewed interest in it.
It is well known that, if one normalizes the amplitude of
fluctuations to
the COBE DMR signal (Smoot \etal 1992),
then the standard cold dark matter model (CDM)
produces too high a small-scale
velocity dispersion (Davis \etal 1992).
There are other problems due to
the shape of
the power spectrum which are independent of amplitude normalization.
A recent review of the triumphs and defects
of the standard CDM scenario is presented in Ostriker (1993).
The mixed dark matter model
was proposed as an interesting alternative
to the CDM model,
which should produce a better agreement with observed
small-scale velocity dispersion measurements
and other observational constraints.
The physical basis for believing in the plausibility
of this approach (two species of non-interecting particles)
is presented in DSS and KHPR.

The rest of the paper is organized in the following manner.
Section 2 gives a brief description of
the equations used
[for a detailed description of equations and
numerical techniques used, see Cen (1992)];
\S 3 briefly describes the method to set up the initial conditions
[see also Cen (1992) for a detailed description of
the procedure to set up the initial conditions];
\S 4 gives the results of the simulations;
\S 5 assembles our conclusions.
\bigskip
\medskip
\centerline{2. EQUATIONS AND NUMERICAL TECHNIQUES}
\bigskip
\medskip
There are two sets of equations,
one for the baryonic fluid and the other for
collisionless dark matter particles.
For the baryonic fluid we have
eight time dependent equations as follows:
the mass conservation equation of total baryonic matter,
the three ionization rate equations of H I, He I and He II,
the three momentum equations in three directions and
the energy equation.
Locally,
we also satisfy charge conservation and the gas equation of state:
$P=n_{tot}kT$.
The set of equations for the collisionless dark matter
particles consists of three equations for change of momentum  and
three for change of position.
In addition, we have the equation relating
the density field to the gravitational forces, i.e.,
Poisson's equation for the perturbed density, and
the two Einstein equations for the evolution of the cosmic comoving frame.
    Details of all the equations are presented in Cen (1992).

    The UV/X-ray radiation field (as a function of frequency and time) is
    calculated in a spatially averaged fashion.
    Changes in other quantities are computed each time step in each cell.
    Ionization, heating and cooling, are computed in a detailed
    non-LTE fashion.

 In terms of numerical technique,
 the dark matter evolution is computed with a standard PM code.
 The dark matter density and the gravitational forces
 exerted on dark matter particles are found using the Cloud-In-Cell
 (``CIC'') algorithm [\cf Hockney \& Eastwood (1981);
 Efstathiou \etal (1985)].
    The gravitational potential, due to both baryons and dark matter,
    is calculated by solving Poisson's equation with periodic
    boundary conditions utilizing an efficient FFT algorithm.

\bigskip
\centerline{3. INITIAL CONDITIONS}
\bigskip
We adopt the analytic fitting formulae for
initial power spectrum transfer functions for
both cold dark matter particles and hot dark matter particles
given in KHPR.
The initial power spectrum transfer function for the baryonic
matter is assumed to follow that of the cold dark matter.

The normalization adopted here is $\sigma_8=0.67$ as in KHPR, which
is fixed by the COBE DMR signals.
The ratio of cold to hot matter also is taken from KHPR.
Standard light element nucleosynthesis (Walker \etal 1991)
determines $\Omega_b$
with our choice of $H$ ($\Omega_b=0.06$ at
the upper end of the permitted range) .
The initial realization of each simulation
is generated by assuming that
the phases of the waves are random and uncorrelated.
The initial dark matter density field and baryon density
field are generated using the same phase information,
although the amplitude of the corresponding modes are different
due to the different power spectra.
The initial peculiar velocity field is then obtained
by the Zeldovich approximation
(\cf Zeldovich 1970).
However, since the hot dark matter component
has a non-trivial random velocity component,
we try to model this velocity component for hot particles
by adding in quadrature the random velocity drawn from
a Fermi-Dirac distribution (following KHPR)
to each pair of hot particles (with the same amplitude but opposite
directions).
To summarize our adopted parameters are as follows:
$h=0.5$,
$\Omega_c=0.64$,
$\Omega_h=0.30$,
$\Omega_b=0.06$ and $\sigma_8=0.67$,
the same parameters as found to be best in DSS, RT and KHPR.

After we have made the simulations, a small error in the initial
power spectrum and an error in the treatment
of the initial velocity generation were brought to our attention
by KHPR. But fortunately, the two errors happen
to compensate one another and the net effect is small
(the rms error for position is $11h^{-1}$kpc
and the rms error for velocity 1.6\%).

\bigskip
\medskip
\centerline{4. RESULTS}
\bigskip
\medskip
\centerline{\it 4.1 Hydrodynamic Simulations}
\medskip
\centerline{\it 4.1.1 Temperature and Density}
\medskip
Four different models are computed with
box sizes of $L=(64, 16, 4, 1)h^{-1}$Mpc, respectively.
We use $128^3$ cells with $128^3$ cold dark matter particles
and
$2\times 128^3$ hot dark matter particles
in each of these simulations.
Thus the nominal resolution in the four
simulations ranges from $500h^{-1}{\rm kpc}$ in the largest box to
$7.8h^{-1}{\rm kpc}$ in the smallest box with actual resolution in the
hydro code about a factor of $2.5$ worse than this.
While resolution of the code is insufficient to answer
many questions of interest, the {\it comparison}
between the results found here and those
presented in CO92 should be very instructive.
In that paper we examined the standard CDM scenario
with the normalization $\sigma_8=0.67$
which is the same as that adopted here.

The larger scale simulations suffer most from the defects of insufficient
resolution, the smaller
scale simulations from the lack
of non-linear long waves which would be truly present
if we had a larger box.
Thus, in the largest box we know that we are underestimating
cooling and condensation of self gravitating small-scale objects,
whereas in the smallest box the omission of long waves makes
the simulation not correct on average in that the gas will
be less violently shaken and thus cooler than the average piece
of the universe at that scale.
Since temperatures are underestimated,
the rate of condensation of self
gravitating objects is overestimated in this small box
(as compared to the average).
The reason for this is that
the Jeans mass at $10^4$\Kel,
where cooling decreases rapidly,
is typically larger than the cell mass.
We do not attempt to model specially those overdense regions where galaxies
preferentially form.
It is likely that our small boxes
(since they have a density equal to the cosmic mean)
underestimate the rate of galaxy formation in regions of high
density, but they {\it overestimate}
galaxy formation as compared to the average cosmic volume of that size.
In other words, the small box
is not a fair sample of the cosmic volume of that
size; it would be necessary to perform many independent simulations with
varying mean density (averaged over the box) to overcome
this weakness to some extent.

The four simulations were run in the following order.
First, we ran the $L=16h^{-1}$Mpc box simulation,
since most of the radiation which is important for
ionizing hydrogen and helium
comes from the scales contained in this box
according to our previous tests.
Second, we ran the $L=64h^{-1}$Mpc box simulation with input radiation
emissivity obtained from the $L=16h^{-1}$Mpc run.
This, we consider the simulation providing the
most useful results.
Third, we ran the $L=4h^{-1}$Mpc box simulation with input radiation
emissivities obtained from both the $L=16h^{-1}$Mpc and $L=64h^{-1}$Mpc runs.
Last, we ran the $L=1h^{-1}$Mpc box simulation with input radiation
emissivities obtained from all three larger box runs.
The reason we ran our models in the given order is
presented in earlier papers of this series.

All the runs started at $z=20$.
As noted, the smaller boxes $L=(4,1)h^{-1}$Mpc are useful only
in a limited sense.
After waves longer than the box size become
nonlinear, calculations on these small scales have little validity.
Besides, in this MDM model, the hot neutrino component has a thermal
motion which is too large to be captured by small-scale $\le 1h^{-1}$
potential wells even at late times;
therefore, in the smaller boxes the missing
long waves should have a larger effect than in the CDM model case,
were they present.
But these simulations provide useful information nonetheless.
The large-scale (missing waves) would have heated the gas on smaller
scales to higher temperatures than obtained when this long
wavelength power is missing.
Thus, formation of cooled, bound objects on these small scales
would have been {\it less}
in a computation with still larger dynamic range
than we have calculated in this paper.
Since one of our main points (already seen in other models)
is that most of the mass does not
fragment into tiny lumps in the MDM picture,
this point is strengthened by our
inclusion of the small-scale boxes,
even if they only
allow an upper bound to be put on the amount of mass
in isolated small-scale structures.

Figure (1) shows the actual initial (scaled to $z=0$)
and final power spectra of the four simulations.
The initial baryonic power spectra are assumed
to be the same as that of the cold dark matter.
The vertical
thick solid label shows the place in the spectrum at $8h^{-1}$Mpc which
we use to parameterize the amplitude of the spectrum.
We see that in the final state, at $z=0$,
on scales $\lambda>2h^{-1}$Mpc hot dark matter
component acts just like the cold dark matter component,
i.e., clusterings of both components on scales $r\ge 1h^{-1}$Mpc
are similar even though the initial spectra are quite different from
one another.
On the other hand, as expected, the hot dark matter component clusters less
on smaller scales due to its thermal motion.
One important point to notice is that any
simulation with box size $\le 10h^{-1}$Mpc
(including the two smaller boxes in this paper)
significantly underestimates the hot dark matter component
density fluctuations.

Now let us turn to the results obtained.
We will compare throughout with the standard CDM run we made (CO92),
which has exactly the same normalization $\sigma_8=0.67$.
On the intermediate $8h^{-1}$Mpc scale
the (integrated) amplitudes are almost identical.
This model has more large-scale power and less small-scale power than
the CO92 run.
The upper panels of Figures (2a,b,c,d)
show the evolution of mean volume-weighted (solid lines),
and mass-weighted (dotted lines) temperatures
as a function of redshift.
Heavy lines show this work and light lines show CO92 run.
Also shown is
the corresponding mean proper peculiar kinetic
energy density (dashed lines) in units of Kelvin.
The simulations are displayed
in the following order:
(a) $L=64h^{-1}$Mpc,
(b) $L=16h^{-1}$Mpc,
(c) $L=4h^{-1}$Mpc,
(d) $L=1h^{-1}$Mpc
(this order is maintained in subsequent figures.)

We see that, in the simulation
with box size $L=64h^{-1}$Mpc [Fig (2a)],
the final mean mass-weighted temperature exceeds $2\times 10^6$\Kel
representing the small fraction of strongly shock heated gas in regions
like the Coma cluster of galaxies.
Similar results are found in CO92 for the standard CDM model
with the same $\sigma_8$ normalization, which is expected since
hot neutrinos behave essentially the same as cold dark matter
on these scales.
In the smallest boxes, $L=(4,1)h^{-1}$Mpc,
the mean temperatures stay at about $10^4$\Kel because
cooling processes (mainly the hydrogen and helium collisional excitation
cooling) are important and the cooling time is short
compared with the Hubble time, so baryonic matter can be
shocked and then cool to remain at these temperatures.
Besides, the shocks on these smaller scales are weaker
(due to omission of waves larger than its box size,
some of which would have entered the nonlinear regime at $z=0$,
were they present)
compared with those in the bigger boxes, where shock
heats baryonic matter to temperatures $\ge 10^6$\Kel.

One main difference which we found for the smaller boxes in this model
compared to the CDM model
is that the temperatures are much lower here, presumably due
to non-clustering nature of still hot neutrinos on these scales.
This difference is most noticeable in the smallest ($L=1h^{-1}$Mpc) box
[Fig (2d)].

The lower panels of Figures (2a,b,c,d)
show the evolution of the variances of
the baryonic and dark matter density on the scale of the cell size of
each simulation, which are defined as
$$\eqalignno{\sigma_M^2&\equiv <\hskip -4pt\rho^2\hskip -4pt >\hskip -4pt
/\hskip -4pt <\hskip -4pt\rho\hskip -4pt >^2-1\hskip 0.4cm ,&(\the\eqtno
)\cr}$$
\advance\eqtno by 1
where $M=(d,b)$ and $\sigma_d$ is the dark matter density variance,
$\sigma_b$ is the baryonic matter density  variance.
In the bigger boxes the dotted line (dark matter) shows higher fluctuations.
In the smaller boxes the gas component (solid line) has higher fluctuations.
We define a bias factor as follows:
    $$\eqalignno{b(\triangle l) &\equiv {\sigma_{b}(\triangle l)\over
\sigma_d(\triangle l)} \hskip 0.4cm .&(\the\eqtno )\cr}$$
\count77 =\the\eqtno
\advance\eqtno by 1
Again heavy lines are from this work and light lines
from CO92.
We find that on scales less than $0.125h^{-1}$Mpc,
cooling processes are important, which leads to the ``biased''
formation of overdense
baryonic objects:
baryonic matter is more clumpy than dark matter on these scales.
But for larger scales, cooling processes
are not significant enough at later times to play an
important role.
Therefore,
we find that on scales larger than $0.125h^{-1}$Mpc, the situation is
just the opposite, i.e., dark matter is more clumpy
than baryonic matter.
Note that this is different from saying that the galaxy
distribution follows (or does not follow)
the mass distribution, since the
baryonic mass distribution
is significantly different from the galaxy distribution
(\cf Katz, Hernquist \& Weinberg 1992;
Cen \& Ostriker 1992c).
Similar results were found in the study of
the standard CDM model (CO92),
in the tilted CDM (TCDM) model (\cf Cen \& Ostriker 1993a)
and
in the CDM$+\lambda$ model (\cf Cen, Gnedin \& Ostriker 1993).
As was stressed before, less cooling in the big simulation
box underestimated the galaxy formation rate.
But quantities such as temperature and Zel'dovich-Sunyaev effect
are only affected weakly since most of the energy which
eventually turns into heat is present in the simulation box;
small-scale waves do not significantly contribute to the
entropy generation although their effect on cooling
in dense regions is, no doubt, very important.

Comparison with the standard CDM result (CO92)
using the identical code (and amplitude, $\sigma_8$)
is instructive.
The density fluctuations in the MDM models
are considerably smaller than those in the CDM models
with the same box sizes, with the differences being
larger for smaller boxes.
This is again due to the fact that neutrinos are hot
enough to escape small-scale potential wells even at $z=0$.
Quantitatively, we find that in the two bigger boxes $\sigma_M$
for the gas reaches $2$ [i.e., $(\delta\rho/\rho)_{rms}=2.3$]
at $z=(0.6,1.4)$ whereas it was $z=(1.0,3.0)$ in the standard CDM  run
having $\sigma_8=0.67$. In the two smaller boxes,
which give a better indication of the initiation of galaxy formation,
we look for the epoch when $\sigma_M(gas)=10$,
and find $z_{10}=(0.7,1.0)$, whereas
in the CDM run this same level of nonlinearity
occured much earlier at
$z_{10}=(2.3,3.4)$.

Figures (3a,b,c,d) show the volume-weighted
histograms of temperatures of cells at four epochs.
Figures (4a,b,c,d) show the mass-weighted
histograms of temperatures of cells at the same epochs.
Comparing these figures with the same numbered figure in CO92
we note the temperature distributions are similar
for the largest box ($L=64h^{-1}$Mpc) in the two cases (MDM vs CDM),
while there is a trend of less hot gas in the MDM model
than in the CDM model for smaller boxes.
This is,
once again,
due to the fact that the MDM model
has less small-scale power
than the CDM model.

Figures (5a,b,c,d) show some typical slices with contours of
baryonic matter density, total dark matter density
(cold + hot components)
and baryonic matter temperature at $z=(2,0)$.
Notice that linear structures are more visible in the gas
than in the dark matter.
These structures arise from the intersection of sheets
(``pancakes") within our displayed slices.
In the dark matter we expect that
perturbations with $\vec K$
vectors within the pancakes will be relatively unstable.
Note also that in Figure (5a),
which shows matter on a large-scale,
the dark matter is more concentrated than the baryons,
whereas in Figures (5c,d), on small scales,
the baryons are more clumped.
A somewhat more filamentary appearance is seen in the gaseous structures
in the MDM model than in the analogous CDM model,
a feature expected from the work of Zel'dovich,
Doroskevich and colleagues
(Shandarin, Doroshkevich, \&  Zeldovich 1983).

In Figures (6a-6d) we contrast the structures
seen in hot and cold dark matter particles.
On the largest scale (6a) differences are not apparent
to the eye, but in the smallest scale box the
fluctuations in the HDM component are grossly
smaller than those of the CDM component,
even though both feel the same gravitational potential.
This effect would clearly have the virture of providing an
increasing mass to light ratio with increasing scale, a trend
noticed from early investigations of the
subject.

Figures (7a,b,c,d) show the ($\rho, T$) contour plots.
The innermost contours represent
conditions of
density and temperature in the most common cells at $z=0$.
Comparison with standard CDM model
is very informative.
Note that in all three smaller boxes there is
a distinctive feature, i.e.,
certain regions with
high densities ($\rho/\bar\rho > 10^2$)
but relatively low temperatures ($\sim 10^4$\Kel)
where the gas is near the peak of the cooling curve.
We found the same feature in standard CDM model (CO92)
but the largest densities (with low tempeatures)
are smaller in this MDM case than in the CDM case.
Also, there are
regions having both high densities and high temperatures
in Figure (7a) representing the
hot X-ray emitting gas in the great clusters.
On all scales the most common cells have a density
$\rho/\bar\rho\simeq 0.1$ and are in ``voids".

\medskip
\centerline{\it 4.1.2 Volume and Mass Distributions}
We now analyze the simulations in another quantitative way.
The baryonic matter in each simulation box
is divided into four components:
(1) virialized, bound, hot objects, which
on the large-scales represent the gas in clusters
of galaxies and on the small scales represent
the L$_\alpha$ clouds --- ``Virialized Gas'';
(2) bound, cooled objects, i.e., collapsed compact objects --- ``Galaxies'';
(3) unbound, hot regions with temperature $\ge 10^5$\Kel --- ``Hot IGM '';
(4) other regions, primarily --- ``Voids''.
The break point at $10^5$\Kel is adopted because it is past the peak of the
``cooling curve".
The quantitative definitions of these regions
are given in CO92.

These four components make a complete set of possible objects and
each cell is classified accordingly.
In Tables (1 - 3) we list the volume and mass weighted
fraction of these four components at six
epochs, $z=(10, 5, 3, 2, 1, 0)$,
for the three different runs with box sizes, $L=64,16,4h^{-1}$Mpc.
In the preceding tables
we do not treat ``galaxy formation'' as irreversible,
which it would be were a true stellar
component to be formed.
Thus, fewer cells satisfy our
criterion to be ``galaxies'' after $z=1$ than at that epoch for Table 1.
The result for the $L=1h^{-1}$Mpc is not tabulated because
we find that all the cells belong to ``Voids".

In the $L=64h^{-1}$Mpc box, in comparison with the CDM simulation,
a larger fraction of the mass is in voids, and ``galaxy formation"
is slightly later and less vigorous.
And in all the boxes galaxy formation is later
in MDM model than in the standard CDM model.
This comparison is also shown in Figure (8).
The most important difference is
that galaxy formation is later in the MDM model,
especially at earlier times,
than in the CDM model with
the same $\sigma_8$.
For example, at $z=3$, the galaxy formation rate is,
by an order of magnitude, lower in the MDM model than in the CDM model,
worsening the already difficult situation of high
redshift quasar formation in the (biased) CDM model.
KHPR, addressing this problem in the context of the approximate
Press-Schechter formalism,
conclude that the model is marginally consistent with the observed
existence of high redshift quasars.
Haehnelt (1993), examining a broader observational data base
and taking a somewhat more critical attitude,
argues that the model considered
in this paper ($\Omega_H=0.3$) (and by the other quoted authors)
is {\it inconsistent} with observations of high redshift quasars.

Now let us look at the properties of the typical collapsed objects
of the four boxes at redshift $z=1$.
The results are shown in Table (4).
We note that the mass weighted mass function
(i.e., mass fraction of collapsed objects)
has a peak around
$<\hskip -5pt m\hskip -5pt>=10^{9.3} M_\odot$.
But we think that this peak would be shifted to a
still larger mass scale
were long missing waves in the smaller boxes included as they
would have heated up the gas medium and stablized
instabilities on small scales.
In a better calculation, with all the longer waves included,
the collapsed fraction would clearly peak at a still larger mass scale
than shown in Table (4),
since the temperature and hence Jeans mass would be higher.
In addition, extra energy input from star formation,
were it included, would also increase the temperature and further
stablize small-scale perturbations.
On all scales the largest fraction of the mass is
in the IGM with
about $2/3$ in the ``Voids" ($T<10^5$K)
and about $1/3$ in the ``Hot IGM" ($T\ge 10^5$K).
A principle difference between MDM model and standard CDM model
is that in CDM model (CO92) we found
a slightly larger mass fraction in the
hot (``Hot IGM") component than that of ``Voids".
Here most of the baryonic mass (as well, of course, as most of the
volume) is in the voids.
In addition, and more significantly,
the galaxy fraction is much less in the MDM picture
as can be seen by comparing Table 5 and with the same
numbered table of CO92.

\medskip
\medskip
\centerline{\it 4.1.3 X-ray Background Radiation}
\medskip
We have calculated the mean
UV/X-ray background radiation field
as a function of frequency as well as
time including absorption by hydrogen and helium
and both free-free and free-bound emission processes.
Figure (9) shows the results at six epochs,
$z=5$ (solid line),
$z=3$ (dotted line),
$z=2$ (short, dashed line),
$z=1.5$ (long, dashed line)
$z=0.5$ (dotted, short-dashed line) and
$z=0$ (dotted, long-dashed line).
Emissivities from both
$L=64h^{-1}$Mpc and $L=16h^{-1}$Mpc runs are included.
The box in the middle shows the observational data by Wu \etal (1991).
We see that the computed MDM model fails by
a factor of $50$ to produce the observed soft X-ray
(0.2 to 1Kev range) background.
The deficit at harder X-rays (1 to 10 KeV range)
is even larger (note that at the high
frequency end the computed spectrum has a very steep slope).
There are two correction terms which need to be taken into account.
First, much of the background is in fact produced
by identifiable AGN sources.
We assume here that approximately half of the
X-ray background radiation is due to discrete AGN sources.
Second, for purely numerical reasons, we know that with the same input
parameters, a better treatment with larger N, larger $L_{max}$ and
smaller $L_{min}$, would increase the X-ray output.
A factor of $3$ increase was found at $1KeV$ in the tests
we made in Cen (1992) for a $128^3$ CDM run with $L=64h^{-1}$Mpc,
$h=0.5$, $b=1.0$, $\Omega=1$ and $\Omega_b=0.1$.
Combining these two factors indicates that
this MDM can make a small but
non-trivial $12\%$ of the residual
soft X-ray background radiation field,
approximately $1/4$
the fraction of the CDM model having the same value of $\sigma_8$.
This is an improvement over the COBE normalized standard CDM model
which overproduces both the X-ray background and correspondingly the number
of high luminosity X-ray clusters (Frenk \etal 1990;
Kang \etal 1994; Bryan \etal 1994).

The strong edges seen in the spectra at $13.6$ eV are due to absorption
by neutral hydrogen.
Meantime, the edges at the $54.4$eV absorption edge
due to once-ionized helium is less
significant simply because there is much less of this species.
The edge at the ionization potential of neutral helium, $24.6$eV,
is seen  at early epochs, but is smaller
because the $24.6$eV edge is too close to the
$L_\alpha$ $13.6$eV edge to be very noticable at our displayed resolution
given the redshift smearing.
At $z=0$, hydrogen and helium are still not completely ionized,
the troughs all remain. Again one should be reminded that
energy feedback (e.g., UV and supernova processes)
from star formation was not included in the simulations;
the effects of these processes will be
smaller than in the CDM model having the same value of $\sigma_8$.

We have computed, but not shown, figures
for the ionization state and opacity [Gunn-Peterson (1965) effect]
for this model.
Needless to say,
without UV from star formation and
supenova energy input into the IGM from young galaxies,
the MDM model is far from satisfying
the Gunn-Peterson test of the high redshift quasar observations,
i.e., the IGM cannot be ionized
by means of shock heating, bremsstrahlung and free-bound radiation.
However, given the nature of late galaxy formation
seen in the MDM model (Fig. 8), we think that
radiation from star formation is not likely to eliminate
this discrepancy.
The reasoning is comparative:
$b=1.3$, CDM with UV input from galaxies is barely satisfactory
(Cen \& Ostriker 1993b) at $z=4$
(\cf Figure 8 of the above referenced paper) and galaxy
formation is lower by approximately $10^2$ in the MDM model.
At $z=5$ we find that less than $10^{-5}$ of the baryons
will have collapsed to possibly form galaxies.
High mass stars with a normal mass function burn
$<10^{-2}$ of the mass with an efficiency
of $10^{-2.5}$ into ionizing photons.
Propagating these parameters through the ionization equations
Miralda-Escude
\& Ostriker (1990) conclude that a collapsed fraction of $10^{-4}$
was marginally satisfactory to satisfy observed
Gunn-Peterson limits and that
$10^{-5}$ would marginally fail.
Tagmark \& Silk (1993) come to a similar
conclusion concerning very late ionization in the MDM scenario.

\medskip
\medskip
\centerline{\it 4.1.4 Zeldovich-Sunyaev Effect}
\medskip
Now we turn to the results of the directly computed
mean Zeldovich-Sunyaev $y$ parameter
at six epochs, $z=(5, 3, 2, 1, 0.5, 0)$ shown in Table (5).
Note that
$\sim 95\%$ ($90\%$ in standard CDM model)
of the contribution to the Zeldovich-Sunyaev effect
comes from the epochs between $z=1$ to $z=0$.
Also the final ($z=0$)
$y$ parameter has a lower value in MDM model than in CDM model.
The reason for this difference is
due to the fact that
small scales waves ($\lambda < 16h^{-1}Mpc$,
which contribute most to entropy generation)
enter nonlinear regime later in the MDM model
than in the CDM model.
Let us emphasize that
Table (5) shows the
{\it directly} computed value of the $y$ parameter.
Due to our inevitable numerical inadequacies we have underestimated $\bar y$.
Using the extrapolation formula derived
in Cen (1992) (\cf equation (76) of that paper),
if we had included all the waves and had
infinite resolution in the calculation,
we would have obtained the following extrapolated value and
an estimated fluctuation of $y$ at $z=0$
when $(N^{-1}, L_{max}^{-1}, L_{min}) \rightarrow 0$,
$\bar y=(5.4\pm 2.7)\times 10^{-7}$,
$\bar{\delta y}=(6.0\pm 3.0)\times 10^{-7}$ on arc minute scales
(where the $\pm$ indicates our estimate of the error of our
extrapolation procedure).
If the reader distrusts our extrapolation procedure,
then Table (5) can be taken as a firm lower bound on $\bar y$ for the
adopted model.

\medskip
\medskip
\centerline{\it 4.1.5 Galaxy and Dark Matter Correlation Functions}
\medskip
A cell belonging to the second
category defined in \S 4.1.2 is called a galaxy.
Further, such cells, if adjacent,
are grouped into a single ``isolated galaxy",
although at our resolution we can not tell the difference between galaxies
and small groups such as the Local Group.
We have found $1502$ such ``isolated galaxies" at $z=1$
in the $L=64h^{-1}$Mpc box.
The reason we identify the galaxies at $z=1$ instead of $z=0$
is for the convenience of comparison with equivalent CDM simulation,
where galaxy formation strongly peaks at $z=1$,
since the breaking of long waves at later times heats up the
baryonic matter causing evaporation of earlier identified galaxies.
In the present model there is still
a similar mass fraction of galaxies at $z=0$ compared to $z=1$
[i.e. less ``evaporation" \cf Figure (7)]
due to weaker shocking in this model.
In a more realistic calculation the transition
to collisionless (stellar) material would be irreversible.
We also randomly selected $2900$ dark matter particles
over the whole box ($L=64h^{-1}$Mpc box) at $z=1$,
which is a good approximation for the representation
of the total mass distribution.

Figure (10) shows the galaxy-galaxy (open circles)
as well as cold dark matter particle-particle (solid dots)
two-point correlation functions
in the simulation with box size $L=64h^{-1}$Mpc at $z=1$.
The errorbars are one sigma Poisson fluctuations.
Also shown is $\xi(R)=(R/5h^{-1}\hbox{Mpc})^{-1.8}$ (dotted line),
the observational data for galaxies (\cf Davis \& Peebles 1983)
scaled down by a factor of $1/2^2$.
This factor is the linear growth factor from $z=1$ to $z=0$ in this model.
It shows that the galaxy distribution is strongly biased
over the mass distribution at this epoch
with the bias factor of about $4$.
The slope of the galaxy-galaxy correlation is
roughly consistent with observed value ($-1.8$) at $z=0$.
Those results are also consistent with those
of KHPR in that the bias needed was found to be approximately $1.9$.
The apparent bias shown by the distance between the open and filled
circles (square root thereof)
is too large by about a factor of $1.7$
but is not trustworthy.
A more precise comparison with observations awaits
a detailed treatment of this scenario (similar to that
given by us for the standard CDM scenarion, Cen \& Ostriker 1993c),
where the galaxy subunits are produced irreversibly
and followed with the PM code.
The reason for a significantly stronger bias in the MDM model
than in the CDM model
is that only fairly deep potential wells are capable of collecting
hot neutrinos causing deepening of the potential wells and hence
inducing galaxy formation.
But the bias is likely to be weaker at $z=0$,
when the neutrinos are cooler.

\medskip
\centerline{\it 4.1.6 Mass Functions and Multiplicity Functions}
\medskip
A cell is called a bound cell if it satisfies the following criteria:
    $$\eqalignno{\phi + 0.5*(v^2+C^2) &< -0.5v_b^2 \quad ,\quad &(\the\eqtno
)\cr}$$
\advance\eqtno by 1
\noindent where $\phi$ is the proper peculiar gravitational potential;
$v$ is the proper peculiar velocity;
$C$ is the local speed of sound.
We take,
as in early papers of this series,
(somewhat arbitrarily)
$v_b^2 \equiv 141^2$(km/s)$^2$;
here $v_b^2$ is the binding energy per unit mass.
We choose such a value of $v_b$ to satisfy
the requirement that
about $70\%$ of the galaxies are in groups/clusters as observations
indicate [Gott \& Turner 1977, Figure (2)].
The definition we have used is arbitrary but corresponds roughly
to what observers identify as
``bound groups".
Of course in an $\Omega=1$ universe {\it all}
galaxies are bound to all other galaxies.
After we have found these bound cells, we group them into a number of
``groups" within each group all the cells are connected (i.e., touching
by at least one side of a cell).
The multiplicity function of the
bound groups, which are defined above
is shown as solid histogram in Figure (11);
also shown (dotted) is the same function for
the $b=1.5$ standard CDM model as dashed histogram.
The difference is statistically significant
for small groups but we do not know which is in better accord
with modern data.

Figure (12) shows the baryonic and total mass/multiplicity functions
of collapsed objects.
Open triangles, filled dots and filled triangles
are collapsed objects from three different boxes,
with box sizes $L=(64, 16, 4)h^{-1}$Mpc, respectively, at $z=1$.
As noted earlier,
the results in the smaller boxes
overestimate the amount of bound material.
The upper panel shows the baryonic mass/multiplicity function
of collapsed objects,
and the dashed line is a fitting formula [equation (5), see below].
The lower panel shows the total mass/multiplicity function
of collapsed objects,
and the dashed line is a fitting formula [equation (6), see below].
    $$\eqalignno{f(M_{bar})dM_{bar}
&=0.06*(M_{bar}/M_{bar}^*)^{-1.3}e^{-M_{bar}/M_{bar}^*}d({M_{bar}\over
M_{bar}^*})\quad, \quad &(\the\eqtno )\cr}$$
\count15 = \the\eqtno
\advance\eqtno by 1
where $M_{bar}$ is the baryonic mass in units of solar mass,
$M_{bar}^*=1.5\times 10^{11}M_\odot$.
    $$\eqalignno{f(M_{tot})dM_{tot}
&=0.01*(M_{tot}/M_{tot}^*)^{-1.3}e^{-M_{tot}/M_{tot}^*}d({M_{tot}\over
M_{tot}^*})\quad, \quad &(\the\eqtno )\cr}$$
\count16 = \the\eqtno
\advance\eqtno by 1
\noindent where $M_{tot}^*=5\times 10^{12}M_\odot$.

Taking the ratio of the fitted number density of simulated galaxies
at $M_{bar}^*$ to the observed number density of galaxies at $L^*$
(Schechter 1976)
gives an estimate of the baryonic mass to blue light ratio,
$(M/L)_1=1.5$.
We obtain a second baryonic
mass to light ratio by matching
the fiducial luminosity
of $L_{B(0)}^*=1.3\times 10^{10}L_\odot$
with $M_{bar}^*$,
which ratio found to be $(M/L)_2=1.5\times 10^{11}/1.3\times 10^{10}=11.5$.
The second estimate is somewhat higher than the first one,
in part due to the low resolution of our simulations.
For example,
we are not able to resolve a system like the Local Group
into separate galaxies.
It is interesting that
these estimates are not grossly inconsistent with one another
and both are not far from what
is obtained in the Galactic disc via the Oort limit
or in globular cluster with the virial theorem.
If we take
the geometric average
of these two estimates, $(M/L)=4.2$,
inserting this derived mass-to-light ratio into
Schechter's original formula yields:
    $$\eqalignno{\phi(M)dM &=0.04({M\over M^*})^{-1.24} e^{-M/M^*}d({M\over
M^*})\quad, \quad &(\the\eqtno )\cr}$$
\advance\eqtno by 1
\noindent where $M^*=5.4\times 10^{10}M_\odot$.
This is shown as the
solid line in upper panel of Figure (12).
In so far as the solid line fits the computations, we can say
that the derived mass functions for collapsed baryonic
matter are consistent with observations
when a baryonic mass to light ratio of
$4-5$ is adopted.
The dashed line in the lower panel corresponds to a Schechter
fit with $M_{tot}/L_B=380$ similar
to the value found observationally in clusters
of galaxies (Trimble 1987).

\bigskip
\centerline{\it 4.2 A Very Large PM Simulation}
\medskip
In order to study statistical properties of clusters of galaxies
as well as those of galaxies on large-scales,
a larger simulation volume is desired.
Our hydrodynamic simulations,
although providing much more detailed physical treatment,
do not have a large enough volume for this purpose.
Besides, a collisionless PM approach should be valid
on very large-scales where thermodynamic processes play
a much less important role than on smaller scales.
We made one large PM simulation with a box size of
$320h^{-1}$Mpc,
$200^3=10^{6.9}$ cold dark matter particles
$2\times 200^3=10^{7.2}$ hot dark matter particles
utilizing a $400^3$ mesh.
The resolution of this simulation is $0.8h^{-1}$Mpc,
which is adequate for the study of masses and correlation functions
of rich clusters.
The volume ($14h^{-3}$Mpc$^3$) within the Abell radius
($1.5h^{-1}$Mpc) corresponds to about 28 cells.
This $640$Mpc simulation is to be compared with the
$200$Mpc simulation by KHPR and the $14$Mpc simulation by
DSS.
Our resolution with a $400$ mesh is nominally $1.6$Mpc,
the same as the nominal resolution of KHPR and
far larger (worse) than the resolution
of the small-scale P$^3$M simulation of DSS.
Thus our work should (numerically)
provide the best statistical information about large
scale features (bulk flow, cluster-cluster correlations etc)
and DSS the best information about the small-scale dark matter
distribution.

\centerline{\it 4.2.1 Power Spectrum}
Figure (13) shows the initial (linearly scaled to $z=0$)
final power spectra of the PM simulation.
The thick, solid line is the initial ($z=20$) power spectrum of
the cold dark matter component in the MDM model.
The thick, dashed line is the initial power spectrum of
the hot dark matter component.
The thin, solid line is the final ($z=0$) power spectrum of
the cold dark matter component.
The thin, dashed line is the final power spectrum of
the hot dark matter component.
For the purpose of comparison also shown is
the final (z=0) power spectrum for the COBE-normalized CDM model
(thin dotted line).
We see three things in this figure.
First, the initially ($z=20$)
noticeable difference in the HDM and CDM power
spectra on scales ($\lambda <5h^{-1}Mpc$)
diminishes at $z=0$ (for $\lambda > 2h^{-1}$Mpc)
due in part to the nonlinear evolution and in part to the interactions
(gravitationally)
between CDM and HDM components.
Second, the final power spectra (both MDM and CDM models)
have a slope of $\sim -1$ in the range $\lambda=5-30h^{-1}$Mpc;
this is a purely nonlinear effect.
Finally, the COBE-normalized CDM model has much
higher fluctuations on scales $1-80h^{-1}$Mpc than
has the CDM model,
the largest difference being $2.3$ in amplitude on the scale
$\lambda\sim 6h^{-1}$Mpc.

\medskip
\centerline{\it 4.2.2 Correlation Function}
Figure (14) shows the two-point correlation functions
for cold dark matter particles and hot dark matter particles
separately.
In the left hand panel at $z=0$ we see
that, on scales $\ge 1h^{-1}$Mpc,
cold dark matter and hot dark matter are distributed similarly.
In other words, the initially hot neutrinos have cooled down sufficiently
by $z=0$ that they have fallen into the gravitational potential
wells of cold dark matter.
The right hand panel shows the situation
at $z=2$.
We see that at that epoch a small difference between the two species remained.

\medskip
\centerline{\it 4.2.3 Cluster Properties}
Now we turn to the clusters of galaxies,
which are
the largest known
gravitationally bound systems
in the universe. We here concentrate on three
fundamental observables for clusters of galaxies:
the cluster-cluster two point correlation function,
the cluster mass function and the cluster merging rate.
For this set of issues our PM simulation should
have significant advantage over prior work on the MDM scenario.

We select the clusters using
an adaptive friends-of-friends linking algorithm.
Then we determine the linking length $b_{ij}$
between the $i$-th and $j$-th particles by
$
b_{ij} = {\rm Min}[{L_{\rm box}/ N^{1/3}},\beta ({1 \over 2})^{1/3}
({1/n_i(a_s)} +
{1/ n_j(a_s)})^{1/3}],
$
where $L_{\rm box}$ is the box size, $N$ is the total number of
particles in the box,
$n_i(a_s)$ is the local number density at
the $i$-th particle's position smoothed over a gaussian window of $a_s$.
We use $a_s=10h^{-1}$Mpc and $\beta=0.25$.
The linking scheme is not sensitive to $a_s$
(e.g., $a_s=5$ or $10h^{-1}$Mpc yields similar results).
The $\beta$ parameter was selected by testing that the linked
groups are neither considerably smaller than the typical
$1.5h^{-1}$Mpc radius observed for rich clusters (see below)
as would happen for small $\beta$, where only the small dense
cluster core is linked,
nor considerably larger as would occur for too large a $\beta$, where
clusters are linked with other neighboring clusters.
The results are not sensitive to small variations
in the selected $\beta$ (Bahcall \& Cen 1992).
{}From this catalog of grouped objects we select
all clusters above a threshold mass within a sphere of
$1.5h^{-1}$Mpc radius
of the cluster center (for proper
comparison with observations).
This yields a final list of clusters and their $1.5h^{-1}$Mpc masses
at $z=0$.

Figure (15) shows the computed cluster mass
functions for MDM model
as well as that from observations
(Bahcall \& Cen 1993).
We see that this model predicts a cluster mass function
about 4 times higher than observed
(with observed masses determined by virial and X-ray temperature
methods agree well with one another, Lubin \& Bahcall 1993),
which is significant since
the observational uncertainty is about a factor of 2.
KHPR agree in their estimate of the mass function
predicted by the model but
stress the observational uncertainties.
While the observations are certainly incomplete we doubt
that this can account for the discrepancy.
The error is of the same sign but not as large in amplitude
as for the pure CDM model.
A pure HDM model,
normalized to COBE, of course produces
too few clusters,
thus we expect that one could find an MDM model
with $\Omega_{CDM}<0.7$ which would be
satisfactory with regard to
this test.
However, such a model would be worse with regard to the early
formation of structure.

Figure (16) shows
the two point correlation function of Abell $R \geq 1$ clusters
in real space,
with mean separation of $55h^{-1}$Mpc, from our
simulated Abell clusters and from observations (Bahcall 1988)
(the computed correlations on scales $r\le 5h^{-1}$Mpc
is probably underestimated due to our limited numerical resolution
of the cluster identification scheme).
We see
that the cluster correlation in this model
is marginally consistent with observations
and is better than the COBE normalized CDM model
and significantly uncertainty still exists
concerning the observational situation..

Figure (17) shows the merging rate in the MDM models and also
the CDM model ($\sigma_8=0.77$) for comparison.
The measure of merging in Figure 17 is based
on an identification of cluster-like gravitating systems
described above.
By comparing the member particles of each cluster at
$z=0$ with the member particles of each cluster at redshift $z$,
we identify the parent cluster for each present-day
cluster as the cluster at redshift $z$
with the maximum number of overlapping members.
Then the fractional mass change in the cluster is
$$
	\left( {\Delta M\over M}\right) _z = 1 - {M_z\over M_0},
\eqno(\the\eqtno )
$$
\advance\eqtno by 1
for parent and present cluster masses $M_z$ and $M_0$.
We compute this statistic for the most massive clusters in the
simulation, with the lower
mass limit chosen so the comoving number density
is $9.4\times 10^{-6}h^3$Mpc$^{-3}$.
Although the normalizations for the two models
and the final distribution of dark matter
are similar,
there is a much larger cluster merging rate in the MDM model
than in the CDM model.
We believe that the reason is that at later times when neutrinos get
sufficiently cooled down, they start to be collected
at the great clusters.
At early times the potential wells are less deep and
the neutrinos are hotter;
the two effects cooperate to reduce
the effective $\Omega$ for cluster material.

In MDM the median change in the cluster mass is
$(\Delta M/M)_{0.3}=0.52$ from $z=0.3$ to the present,
and $(\Delta M/M)_{1.0}=0.90$ from $z=1$. These are
considerably larger than the corresponding values
$(\Delta M/M)_{0.3,1.0}=0.29$ and 0.77 for CDM.
The rapid merging rate in CDM
is discussed by Frenk {\it et al.} (1990).
We suspect that the evolution of the great
cluster properties will be different enough between MDM and
CDM (\cf also Figure (15)
to provide a meaningful comparative test.

\medskip
\centerline{\it 4.2.4 Velocity Information}
We compute two statistics with regard to the velocity field.
First, in Figure (18a) we show
the one-dimensional relative
velocity dispersion defined as
$$
        v_{1d} = \langle [v_x(1)-v_x(2)]^2\rangle ^{1/2}/\sqrt{3}.
\eqno(\the\eqtno )
$$
\advance\eqtno by 1
This is
averaged over particles, that is, $v_{1d}$ is a mass-weighted statistic.
At $1h^{-1}$Mpc separation the rms value for
the 1d velocity dispersion
is $605\pm 8$km/s.
Correcting this for the velocity bias
that we found on the $1h^{-1}$Mpc scale in Cen \& Ostriker (1992c)
of $0.8\pm 0.1$ (for the very similar $b=1.3$ CDM model)
we find $v_{1d}(gal)=484\pm 6$ which
is to be compared with
$340\pm 40$km/s.
The discrepancy remains but is considerably less than
in the COBE normalized standard CDM model.
Also shown is the data from Davis \& Peebles (1983).
It is seen that this MDM fares similarly as the standard CDM
model with same $\sigma_8$, but in disagreement with observed value.
The physical velocity bias (\cf Cen \& Ostriker 1992c)
of $b_v=0.8$ we see ($1h^{-1}$Mpc)
is not able to bridge the gap.
We have compared our results with these of KHPR
with regard to this all important statistic.
Specifically, Figure (18) can be compared with Figure (10)
of that paper.
Qualitatively the two sets of results show a similar
dependence on $r$ in the range $2$Mpc$< r h < 8$Mpc,
where both calculations might be valid.
But despite the identical assumed power spectra and normalizations,
and very similar numerical methods, our results for $v_{1d}$ are larger
than those in KHPR by about a factor of $1.5$.
The difference is large enough so that KHPR could assert satisfactory
agreement with observations, whereas we find that the
disagreement is probably significant.

What is the truth here?
We believe that the difference is primarily
due to our larger box size ($320h^{-1}$Mpc in our case vs $25h^{-1}$Mpc
in KHPR),
which allows longer waves and more high velocity dispersion clusters,
and due to the fact that in KHPR pairs with velocity difference
greater than $1000$km/s are excluded.
We did the following exercise to test this hypothesis.
We randomly select 100 boxes of size $25h^{-1}$Mpc within
our $320h^{-1}$Mpc
simulation box and computed the above statistic separately
for each of the subboxes.
We then group the results to
show the dependence on the mean density of the subbox being studied.
The results for the 1-d
velocity dispersion at $1h^{-1}$Mpc separation
as a function of mass overdensity of the subboxes
relative to the global mean
are shown (Figure [18b]) for two case: the open circles
with and filled dots without velocity pairs $>1000$km/s.
Also shown are the value from KHPR (thin horizontal
arrow at the left axis)
and our computed value (thick horizontal arrow).
The dashed histogram indicates the distribution
(shown by the right vertical axis)
of the subboxes as a function of their overdensities.
Note that the average pairwise velocity dispersion
(the open circles weighted by the dashed histogram)
is not the same as indicated by the thick solid arrow,
since the former is a volume-sampling and the latter
is a particle-sampling.

By construction, the mean density
of the KHPR $25h^{-1}$Mpc box was unity,
and their result (thin arrow)
is consistent with what we obtained from our
subset
of boxes with
$0.8<\rho/\langle\rho\rangle < 1.2$.
We see that it is not surprising
that KHPR obtained a lower value (by a factor of 1.5)
than ours.
The larger value found in our work
is $605\pm 69$km/s ($1\sigma$ dispersion)
or $\pm 8$km/s (probable error)
due simply to use of a larger box which can include
more long wavelength power.

Next, in Figure (19) we show
the scalar correlation function for
the mass peculiar velocity field defined as
$$
	\psi (r) = \hbox{sign}|\langle
	\vec v(\vec x)\cdot\vec v(\vec x+\vec r)\rangle |
	^{1/2},\eqno(\the\eqtno )
$$
\advance\eqtno by 1
again mass weighted.
The prefactor means $\psi$ is given the sign of the
autocorrelation function.
We see that the coherence length $l_{v}$ (defined as the scale
where this statistic drops to the value half that at
zero separation) is $\sim 40h^{-1}$Mpc in agreement with that
of the standard CDM model, but smaller than
some recent observations which indicate very large-scale
bulk motion (Lauer \& Postman 1992).

\medskip
\centerline{\it 4.2.5 Dipole Issue}
We consider finally the relation between the large-scale
mass distribution and the peculiar velocity of the
Local Group. In linear perturbation theory, the peculiar velocity
at position $\vec r$
produced by the mass distribution represented by point masses
$m_i$ at positions $\vec r_i$ is
$$
	\vec v = {GH_o\Omega ^{0.6}\over 4\pi G\rho _b}
	\sum m_i{\vec r_i -\vec r\over |\vec r_i -\vec r|^3}.
\eqno(\the\eqtno )
$$
\advance\eqtno by 1
The scaling with the density parameter $\Omega$ is a useful
approximation if the cosmological constant vanishes or if the
universe is cosmologically flat (Peebles 1984). In an application
of equation (10) to a catalog of mass markers, the sum must be
truncated at some maximum distance $R$. The truncation causes a
misalignment of the predicted velocity and the observed velocity
$\vec v_{lg}$ of the Local Group relative to the CBR, and, if the
observed and predicted values of $v_{lg}$ are used to estimate
$\Omega$, the missed mass fluctuations beyond the depth of the
catalog can produce a systematic overestimate of the density
parameter (Juszkiewicz, Vittorio, \& Wyse 1990).
We investigated these effects in the MDM model runs by
comparing the prediction of equation (10) when the sum is
truncated at distance $R$ (by a Gaussian window $e^{-r^2/2R^2}$)
to the actual peculiar velocity
computed as the weighted sum
$$
	\vec v\equiv \sum \vec v_iW_i/\sum W_i ,
\eqno(\the\eqtno )
$$
\advance\eqtno by 1
where the $\vec v_i$ are the dark matter particle velocities and the
weight function decreases linearly with distance from the chosen
origin to $W_i=0$ at distance $r=2.5h^{-1}$Mpc.
Equation (11) averages over the small-scale motions, as one does
for the motion of the Local Group, while preserving the velocity field the
scales where we can trust our code.
Figure 20 compares the MDM, CDM and PBI (see Cen, Ostriker, \& Peebles 1993)
distributions of the misalignment angle
$\theta$ between the actual velocity $\vec v$ and
the predicted direction as a function of the limiting distance.
The distribution of $\theta$ is broader in PBI, because the
large-scale density fluctuations are larger.
But MDM and CDM yield similar distributions.

Next, we examine in
Figure 21 the distribution of results of estimating the
mass density by setting the magnitude of the actual
velocity equal to the magnitude of the sum in equation (10)
truncated at distance $R$, and then solving for the apparent density
parameter $\Omega _e$.
We see that it is necessary to study
a very large volume in order to get a reliable estimate for
$\Omega$.
At $R=10h^{-1}$Mpc half of the observers
in the MDM model would think that $\Omega$ was greater than $7$!
We also see that in MDM one would generally overestimate
$\Omega$ on small scales significantly more than in the CDM or PBI cases.
The sharp upturn of $\Omega_e$ as one goes
to smaller scales is very interesting.
The reason is that at present ($z=0$) the relatively cold neutrinos
still have a significant amount of thermal motion which makes
the relatively shallow potential wells incapable of capturing them.
Since we randomly (uniformly in volume) sample the space,
we mainly sample the underdense regions (which
occupy most of the space and where potential wells are shallow);
in these regions, velocities (as well as densities and potentials)
are in large part not induced
by gravity, and therefore the apparent $\Omega_e$
does not represent the mass density on these small scales.
These figures assume perfect data and complete sampling,
which is clearly unatainable in practice.
A more realistic set of assumptions would have further
increased the dispersion.
A similar set of conclusions with regard to
the determination of $H_0$ was made
by Turner, Cen \& Ostriker (1992).

\bigskip
\medskip
\centerline{5. CONCLUSIONS}
\bigskip
\medskip
Our hydrodynamic simulations of the MDM scenario
utilizing different cell sizes and box sizes to cover the
dynamic ranges of interest
are sufficiently accurate, we believe, to allow us to compute,
with reasonable confidence, properties of the gas distribution on
scales larger than $2.5$ cell sizes and to compare
with the standard CDM model computed with the
identical numerical code.
Our large PM simulation complements our hydro simulation
on large-scales.
Our results show that this MDM model,
while normalized to COBE,
appears to fare similarly as the
standard CDM model with the same $\sigma_8$.

(1) Galaxy formation occurs somewhat later
in MDM model than in the standard CDM model with the
same $\sigma_8$.
The galaxy formation fraction is about $0.01\%$ at $z=3$ and
peaks near $z\sim 0.3$ in the
MDM model while in the standard CDM model the peak
is around $z\sim 0.5$.
At redshift $z=4$ the MDM model
has less galaxy formation by a factor of nearly one hundred
than the CDM model having the same value of $\sigma_8$.
Reducing $\Omega_b$ substantially could
increase the power on small scales
to up to $25\%$ but we doubt
that this would suffice to bridge
the gap and it would produce other problems,
since a lower baryon density reduces the cooling rate
and thus inhibits galaxy formation.

(2) The soft X-ray radiation is far below and thus
    consistent with the observations by Wu \etal (1991).
    But it can still make a non-trivial contribution to the observed
    soft X-ray background, approximately
    $12\%$ of the residual (after taking
    into account of the half contribution from discrete sources)
    X-ray background.
The Zeldovich-Sunyaev $y$ parameters
is computed to be
$\bar y=(5.4\pm 2.7)\times 10^{-7}$
with fluctuations
$\bar{\delta y}=(6.0\pm 3.0)\times 10^{-7}$
on arc minute scales, which are
below current observational limits.

(3) With our scheme of identifying galaxy formation candidates at $z=1$
we find that the final, computed bias of galaxy distribution
over mass is $\sim 4$, a value
which is larger than the assumed value ($b=1.5$).
But a more quantitative comparison between simulated galaxies
at $z=0$ with the observations awaits a simulation where
galaxy formation is treated to be irreversible (like the one
for CDM model of Cen \& Ostriker 1993b,c).
The two-point correlation function of galaxies has
approximately the correct slope given our crude scheme of
tagging galaxies.

(4) Using physical criteria for the formation of galaxies from
  cooling gas, we find that approximately the correct total mass
  density of baryons collapses to galaxies and that
  these have approximately the correct mass spectrum.
  Specifically, a reasonable fit to the observed Schecter luminosity
  function is obtained if $M_b/L_B=4$ to give
  $M_{bar}^*=5\times 10^{10}M_\odot$
  and $M_{tot}^*=5\times 10^{12}M_\odot$.
  Thermal energy prevents the smallest scales from being
  most unstable with the result that the mass-weighted
  mass function is expected to decline
  for galaxy masses (in baryons) less than $2\times 10^9M_\odot$.

(5) The small-scale ($\Delta r=1h^{-1}$Mpc)
velocity dispersion is $605\pm 8$km/s, which
might be reduced to $480\pm 6$km/s by a reasonable
physical bias, still somewhat in excess of
the observed value.
The large scale coherence of motion
is almost identical to that in the CDM model.

(6) The cluster mass function is larger by a factor of about 4,
and cluster-cluster 2-point
correlation length slightly lower but
marginally consistent with observations.

Overall, this model does not seem to be more successful than
the standard CDM model with the same value of $\sigma_8$,
but it is far better than standard CDM if both are normalized to COBE.
Critical tests for this model check
on whether or not it will provide enough nonlinear
structure at early times.
On the $10-100$kpc scale higher resolution simulations are required
to see if galaxy formation can begin at an early enough epoch to satisfy
Gunn-Peterson and other constraints.
At the $0.1-10$Mpc scale higher resolution studies are needed
to test if there are enough high central density clusters at moderate
redshift in this picture to provide
the gravitational lenses needed to make
the observed luminous arcs.

While this proposed work is still to be done;
it seems to us that no successful MDM model
can or will be found.
The reason is that observational constraints push the unknown,
ratio $r_C\equiv \Omega_{CDM}/(\Omega_{CDM}+\Omega_{HDM})$
in opposite directions.
We adopted $r_{C}=0.7$, the same as other investigations (KHPR, DSS, TR).
In order to produce early enough formation of quasars (Haehnelt 1993)
or galaxies (this paper) and clusters of galaxies (this paper)
a larger value of $r_{C}$ should probably be adopted.
But in order to match the cluster mass function or
the small-scale velocity dispersion a smaller value of $r_{C}$
is required.
It is easy to see that variations of the Hubble
parameter $h$ will
not be able to overcome these difficulties.
It may be that the observations are pushing us firmly
towards a serious consideration of open $\Omega<1$ models.

The research of R.Y.C. and J.P.O. are supported in part by NASA grant
NAGW-2448 and NSF grant AST91-08103.
We thank M. Davis, A Klypin and J. Primack for useful discussions.
\vfill\eject

\centerline{REFERENCES}
\bigskip
\medskip
\refset
Bahcall, N.A. 1988, ARAA, 26, 631
\smallskip
\refset
Bahcall, N.A., \& Cen, R.Y., 1992, ApJ(Letters), 398, L81
\smallskip
\refset
Bahcall, N.A., \& Cen, R.Y., 1993, ApJ(Letters), 407, L49
\smallskip
\refset
Bryan, G.L, Cen, R.Y., Norman, M.L., Ostriker, J.P,
\& Stone, J.M. 1994, ApJ, in press
\smallskip
\refset
Cen, R.Y. 1992, ApJS, 78, 341
\smallskip
\refset
Cen, R.Y., Gnedin, N.Y., Kofman, L.A, Ostriker, J.P. 1992
        ApJ(Letters), 399, L11
\smallskip
\refset
Cen, R.Y., Gnedin, N.Y., \& Ostriker, J.P. 1993
        ApJ, 417, 387
\smallskip
\refset
Cen, R.Y., Ostriker, J.P., Spergel, D.N., \& Turok N. 1991, ApJ, 383,1
\smallskip
\refset
Cen, R.Y., \& Ostriker, J.P. 1992a, ApJ, 393, 1 (CO92)
\smallskip
\refset
Cen, R.Y., \& Ostriker, J.P. 1992b, ApJ, 399, 331
\smallskip
\refset
Cen, R.Y., \& Ostriker, J.P. 1992c, ApJ(Lett), 399, L113
\smallskip
\refset
Cen, R.Y., \& Ostriker, J.P. 1993a, ApJ, 414, 407
\smallskip
\refset
Cen, R.Y., \& Ostriker, J.P. 1993b, ApJ, 417, 404
\smallskip
\refset
Cen, R.Y., \& Ostriker, J.P. 1993c, ApJ, 417, 415
\smallskip
\refset
Cen, R.Y., Ostriker, J.P., \& Peebles, P.J.E. 1993, ApJ, 415, 423
\smallskip
\refset
Davis, M., \& Peebles, P.J.E. 1983, ApJ, 267, 465
\smallskip
\refset
Davis, M., Summers, F.J., \& Schlegel, D. 1992, Nature, 359, 393 (DSS)
\smallskip
\refset
Davis, M., Efstathiou, G., Frenk, C.S., \& White, S.D.M. 1992,
Nature, 356, 489
\smallskip
\refset
Efstathiou, G., Davis, M., Frenk, C.S., \& White, S.D.M. 1985, ApJS, 57, 241
\smallskip
\refset
Frenk, C.S., White, S.D.M., Efstathiou, G., \& Davis, M. 1990, ApJ, 351, 10
\smallskip
\refset
Gott, J.R., III, Turner, E.L. 1977, ApJ, 216, 357
\smallskip
\refset
Gunn, J.E., \& Peterson, B.A. 1965, ApJ, 142, 1633
\smallskip
\refset
Haehnelt, M. 1993, preprint
\smallskip
\refset
Hockney, R.W., \& Eastwood, J.W. 1981, ``Computer Simulations
Using Particles", McGraw-Hill, New York.
\smallskip
\refset
Juszkiewicz, R., Vittorio, N. \& Wyse, R.F.G. 1990, ApJ, 349, 408
\smallskip
\refset
Kang, H., Cen, R.Y., Ostriker, J.P., \& Ryu, D. 1994, ApJ, in press
\smallskip
\refset
Katz, N., Hernquist, L., \& Weinberg, D.H 1993, ApJ, 399, L109
\smallskip
\refset
Klypin, A., Holtzman, J., Primack, J., \& Regos, E. 1993, ApJ, 416, 1
\smallskip
\refset
Lauer, T., \& Postman, M 1993, in preparation
\smallskip
\refset
Lubin, L.M., \& Bahcall, N.A. 1993, ApJ(Letters), 415, L17
\smallskip
\refset
Miralda-Escude, J. \& Ostriker, J.P. 1990, ApJ, 350, 1
\smallskip
\refset
Ostriker, J.P. 1993, ARAA, 31, 689
\smallskip
\refset
Peebles, P.J.E. 1984, ApJ, 284, 439
\smallskip
\refset
Schechter, P. 1976, ApJ, 203, 297
\smallskip
\refset
Shandarin, S., Doroshkevich, A., Zeldovich, Ya. 1983,
 Sov.Phys.Usp. 26, 46
\smallskip
\refset
Smoot, G.F., \etal 1992, ApJ(Letters), 396, L1
\smallskip
\refset
Sunyaev, R.A., \& Zeldovich, Ya.B. 1970, Ap.\& Sp. Sci, 7, 13
\smallskip
\refset
Sunyaev, R.A., \& Zeldovich, Ya.B. 1972, Astr.Astrophys., 20, 189
\smallskip
\refset
Tagmark, M \& Silk, J. 1993, preprint
\smallskip
\refset
Taylor, A.N., \& Rowan-Robinson, M. 1992, Nature, 359, 396 (TR)
\smallskip
\refset
Trimble, V. 1987, ARAA, 25, 425
\smallskip
\refset
Turner, E.L., Cen, R.Y., \& Ostriker, J.P. 1992, AJ, 103, 1427
\smallskip
\refset
Walker, T.P., Steigman, G., Schramm, D.N., Olive, K.A., \& Kang, H.S.
   1991, ApJ, 376, 51
\smallskip
\refset
Wu, X., Hamilton, T., Helfand, D.J., \& Wang, Q. 1991, 379, 564
\smallskip
\refset
Zeldovich, Ya. 1970, Ast. Ap., 5, 84
\smallskip
\vfill\eject

\centerline{\ \ \ \ \ \ FIGURE CAPTIONS}
\bigskip
\medskip
\hsize=5.25truein
\hoffset=2.45truecm
\item{Fig. 1--}
Figure (1) shows
the initial power spectra for
cold dark matter component (thick solid line)
and hot dark matter component (thick dashed line),
and
the final power spectra for
cold dark matter component (thin solid line)
and hot dark matter component (thin dashed line)
for the four simulation boxes separately.
The baryonic power spectrum is assumed to follow initially
that of the cold dark matter component.
The thick solid label shows the place in the spectrum at $8h^{-1}$Mpc,
which we use to parameterize the amplitude of the spectrum.

\item{Fig. 2--}
The upper panels of Figures (2a,b,c,d)
show the evolution of mean volume-weighted (solid lines),
and mass-weighted (dotted lines) temperatures
as a function of redshift.
Also shown is
the corresponding mean proper peculiar kinetic energy density (dashed lines)
in units of Kelvin.
The lower panels of Figures (2a,b,c,d)
show the density variances of baryonic matter (solid line)
and dark matter (dotted line) [cf equation (1) for definitions].
Heavy lines show this work and light lines show CO92 run.

\item{Fig. 3--}
Figures (3a,b,c,d) show the volume-weighted
histograms of temperatures of cells at several epochs,
at $z=5$ (solid line), $z=2$ (dotted line),
$z=1$ (solid line) and
$z=0$ (dotted line).
The peaks at $T\sim 10^{4.5}$\Kel are mainly due to cooling by
by hydrogen $L_\alpha$ lines.
The temperature rises in the bigger boxes at late times
are due to the ultimate breaking of long waves.

\item{Fig. 4--}
Figures (4a,b,c,d) show the mass-weighted
histograms of temperatures
of cells at several epochs,
at $z=5$ (solid line), $z=2$ (dotted line),
$z=1$ (solid line) and
$z=0$ (dotted line).

\item{Fig. 5--}
Figures (5a,b,c,d) show some typical slices of
baryonic matter density, dark matter
density and baryonic matter temperature contour plots
at $z=2$ and $z=0$ for simulations with $L=64, 16, 4h^{-1}$Mpc,
respectively.
All slices are $21$ cells thick.
The contour levels for densities are as following:
$[1+\sigma(\rho_b)]^{I/2}$ for $L=64h^{-1}$Mpc and $L=16h^{-1}$Mpc boxes,
$[1+\sigma(\rho_b)]^{I/4}$ for $L=4h^{-1}$Mpc and $L=1h^{-1}$Mpc boxes,
with $\sigma(\rho_b)$ the density rms fluctuation in the baryonic matter.
The contour levels for temperature are as following:
$[1+\sigma(T)]^{I/2}$ for $L=64h^{-1}$Mpc and $L=16h^{-1}$Mpc boxes,
$[1+\sigma(T)]^{I/4}$ for $L=4h^{-1}$Mpc and $L=1h^{-1}$Mpc boxes,
with $\sigma(T)$ the rms temperature fluctuation in the baryonic matter;
where $I$ is positive integer,
$\sigma(\rho_b) (z=2) = 0.67, 1.12, 2.11, 3.38$,
$\sigma(\rho_b) (z=0) = 3.52, 8.45, 31.87, 20.76$,
$\sigma(T) (z=2) = 1.75, 2.92, 5.60, 6.92$,
$\sigma(T) (z=0) = 9.89, 16.83, 11.38, 5.05$, respectively,
for $L=64, 16, 4, 1h^{-1}$Mpc boxes.

\item{Fig. 6--}
Figures (6a,b,c,d) show one typical slice of
CDM density and HDM density at three
($z=5,2,0$) redshifts
for simulations with $L=64, 16, 4h^{-1}$Mpc, respectively.
All slices are $21$ cells thick.

\item{Fig. 7--}
Figures (7a,b,c,d) show the ($\rho, T$) contour plots.
The innermost contours represent the highest fraction
of cells in terms of volume with these densities and temperatures at $z=0$.
Note that the smaller boxes have certain regions with high densities
but relatively low temperatures while for bigger boxes this feature
disappears.
The contour levels are defined as follows:
$10^{I/4}$, where I is positive integer, $I=0$ corresponds to
the outermost contour, and contours inside it have gradually increasing $I$.

\item{Fig. 8--}
Figure (8) shows the collapsed galaxy fractions
in four four simulations:
thick, solid curve (MDM with $L=64h^{-1}$Mpc),
thick, dashed curve (MDM with $L=16h^{-1}$Mpc),
thin, solid curve (CDM with $L=64h^{-1}$Mpc),
thin, dashed curve (CDM with $L=16h^{-1}$Mpc).

\item{Fig. 9--}
Figure (9) shows the mean radiation at five epochs,
at
$z=5$ (solid line),
$z=3$ (dotted line),
$z=2$ (short, dashed line),
$z=1$ (long, dashed line)
$z=0.5$ (dotted-short-dashed line) and
$z=0$ (dotted-long-dashed line).
The box in the middle shows the observational data by Wu \etal (1990).

\item{Fig. 10--}
Figure (10) shows the galaxy-galaxy
as well as dark matter particle-particle
two-point correlation functions
in the simulation of $L=64h^{-1}$Mpc box at $z=1$.
Open and filled circles are
the galaxy-galaxy two-point position correlation
and
the dark matter particle-particle two-point position correlation,
respectively,
The errorbars are one sigma Poisson fluctuations.
The dotted line is which is the observational data (\cf Davis and Peebles
1983),
$\xi(R)=(R/5h^{-1}\hbox{Mpc})^{-1.8}$
scaled down by a factor of $1/2^2$
(This factor is the linear growth factor from $z=1$ to $z=0$ in this model).
Note that the simulated galaxies are strongly
biased with regard to the dark matter and the observed galaxies.

\item{Fig. 11--}
Figure (11) shows the number density
of groups as a functions of number of galaxies in the group
identified in the $L=64h^{-1}$Mpc box.
This is related to the multiplicity
function but a quantitative comparison with observations
is not appropriate given our poor spatial resolution.
The corresponding curve for the CDM model (CO92)
is shown as the dashed histogram.
We see that MDM model
predicts a somewhat higher level
of low end groups while at the high end
the two models agree.

\item{Fig. 12--}
Figure (12) shows the baryonic and total mass/multiplicity functions
of collapsed objects.
Open triangles, filled dots, filled triangles and filled squares
are collapsed objects from four different boxes,
with box sizes $L=(64, 16, 4)h^{-1}$Mpc, respectively.
The upper panel
of Figure (12) shows the baryonic mass/multiplicity function
of collapsed objects,
and the dashed line is a fitting formula [equation (\the\count15)],
where $M_{baryon}$ is in units of solar mass,
the solid line is the derived Schechter function if mass to light
ratio is 4.2 [equation (5)].
The lower panel of Figure (12) shows the total mass/multiplicity function
of collapsed objects,
and the dashed line is a fitting formula [equation (\the\count16)].
Data roughly fits the Schechter function if the baryon mass to blue light
ratio is in the range $1.5$ to $7.7$.

\item{Fig. 13--}
Figure (13) shows the initial (linearly scaled to $z=0$)
final power spectra of the PM simulation.
The thick, solid line is the initial ($z=20$) power spectrum of
the cold dark matter component the MDM model.
The thick, dashed line is the initial power spectrum of
the hot dark matter component.
The thin, solid line is the final ($z=0$) power spectrum of
the cold dark matter component.
The thin, dashed line is the final power spectrum of
the hot dark matter component.
For the purpose of comparison also shown is
the final (z=0) power spectrum for the COBE-normalized CDM model
(thin dotted line).

\item{Fig. 14--}
Figure (14a) shows the dark matter particle-particle
two-point correlation functions from the large PM simulation
($L=320h^{-1}$Mpc) at $z=0$.
The two dark matter components (CDM and HDM) are shown separately.
The errorbars are one sigma Poisson fluctuations.
The dotted line is which is the observational data (\cf Davis and Peebles
1983),
$\xi(R)=(R/5h^{-1}\hbox{Mpc})^{-1.8}$.
Figure (14b) shows those at $z=2$.

\item{Fig. 15--}
Figure (15a) shows the computed cluster mass functions for MDM model
as well as that from observations (Bahcall \& Cen 1993).
We believe that the discrepancy is significant.
Also shown is that for the COBE-normalized CDM model.
Figure (15b) shows the computed cluster mass functions for MDM model
at three different epochs ($z=0,0.3,1$).
Note that the masses for the clusters
are within $1.5h^{-1}$Mpc (metric not comoving).

\item{Fig. 16--}
The two point correlation function of Abell $R \geq 1$ clusters
in real space,
with mean separation of $55h^{-1}$Mpc, from our
simulated Abell clusters and from observations (Bahcall 1988)
(the computed correlations on scales $r\le 5h^{-1}$Mpc
is probably underestimated due to our limited numerical resolution
of the cluster identification scheme).
We believe that the discrepancy is significant.
Also shown is that for the COBE-normalized CDM model.

\item{Fig. 17--} A measure of cluster merging.
The figure compares the distribution in the fractional mass
change $\Delta M/M$ (eq. [11]) from redshifts $z=0.3$ and
$z=1$ to the present, for clusters with number density
$9.4\times 10^{-6}h^3$Mpc$^{-3}$, in the MDM and CDM models.

\item{Fig. 18--}
Figure (18a) shows the
one-dimensional mass-weighted scalar relative velocity dispersion
[see equation (8)]
$v_{1d}$ as a function of three-dimensional separation $r$.
Also shown is the data from Davis and Peebles (1983)
(which might reasonably corrected upwards by a factor
of $1.25$ to allow for velocity bias on this scale).
Figure (18b) shows
the 1-d
velocity dispersion at $1h^{-1}$Mpc separation in
100 randomly selected $25h^{-1}$Mpc subboxes
as a function of mass overdensity of the subboxes
relative to the global mean for two case: the open circles
with and filled dots without velocity pairs ($>1000$km/s).
Also shown are the value from KHPR (thin horizontal
arrow at the left axis)
and our computed value (thick horizontal arrow).
The dashed histogram indicates the distribution
(shown by the right vertical axis)
of the subboxes as a function of their overdensities.
Note that the average pairwise velocity dispersion
(the open circles weighted by the dashed histogram)
is not the same as indicated by the thick solid arrow
since the former is a volume-sampling and the latter
is a particle-sampling.

\item{Fig. 19--}
The mass-weighted scalar velocity autocorrelation function (eq.
[9]) as a function of separation.
Note that the velocity coherence length is about
$110h^{-1}$Mpc, significantly larger than in the CDM model.

\item{Fig. 20--}
Frequency distribution
of the angle $\theta$ between the actual peculiar velocity $\vec v$ and
the predicted direction as a function of the limiting distance,
for MDM, CDM and PBI models.

\item{Fig. 21--}
Distribution of the effective density parameter (eq. [10]) as a
function of the limiting distance $R$
for MDM, CDM and PBI.
Note both the large dispersion and the systematic
tendency to overestimate $\Omega$ as compared to the true
values (indicated by arrows).

\vfill\eject

\centerline {{\bf Table 1a.} Summary of
the volume weighted fractions in the $L=64h^{-1}$Mpc box model}
\medskip
\begintable
\hfill Redshift\hfill|
\hfill \quad 10\quad\hfill|
\hfill \quad 5\quad \hfill|
\hfill \quad 3\quad \hfill|
\hfill \quad 2\quad \hfill|
\hfill \quad 1\quad \hfill|
\hfill \quad 0\quad \hfill\cr
\hfill \quad \hbox{(1) ``Virialized Gas''}\quad \hfill|
\hfill 0. \hfill|
\hfill 0. \hfill|
\hfill $0.000000$ \hfill|
\hfill $0.000000$ \hfill|
\hfill $0.000004$ \hfill|
\hfill $0.000009$ \hfill\cr
\hfill \quad \hbox{(2) ``Galaxies''\hskip 0.9cm}\quad \hfill|
\hfill 0. \hfill|
\hfill 0. \hfill|
\hfill $0.000005$ \hfill|
\hfill $0.000062$ \hfill|
\hfill $0.000851$ \hfill|
\hfill $0.000456$ \hfill\cr
\hfill \quad \hbox{(3) ``Hot IGM''\hskip 0.8cm}\quad \hfill|
\hfill 0. \hfill|
\hfill 0. \hfill|
\hfill $0.000025$ \hfill|
\hfill $0.000217$ \hfill|
\hfill $0.004745$ \hfill|
\hfill $0.063292$ \hfill\cr
\hfill \quad \hbox{(4) ``Voids''\hskip 1.35cm}\quad \hfill|
\hfill 1. \hfill|
\hfill 1. \hfill|
\hfill $0.999969$ \hfill|
\hfill $0.999721$ \hfill|
\hfill $0.994267$ \hfill|
\hfill $0.936243$ \hfill
\endtable
\bigskip
\bigskip
\bigskip
\bigskip
\bigskip

\centerline {{\bf Table 1b.} Summary of
the mass weighted fractions in the $L=64h^{-1}$Mpc box model}
\medskip
\begintable
\hfill Redshift \hfill|
\hfill \quad 10\quad\hfill|
\hfill \quad 5\quad \hfill|
\hfill \quad 3\quad \hfill|
\hfill \quad 2\quad \hfill|
\hfill \quad 1 \hfill|
\hfill \quad 0 \quad \hfill\cr
\hfill \quad \hbox{(1) ``Virialized Gas''}\quad \hfill|
\hfill 0. \hfill|
\hfill 0. \hfill|
\hfill $0.000000$ \hfill|
\hfill $0.000000$ \hfill|
\hfill $0.000049$ \hfill|
\hfill $0.000064$ \hfill\cr
\hfill \quad \hbox{(2) ``Galaxies''\hskip 0.9cm}\quad \hfill|
\hfill 0. \hfill|
\hfill 0. \hfill|
\hfill $0.000057$ \hfill|
\hfill $0.000799$ \hfill|
\hfill $0.012183$ \hfill|
\hfill $0.005024$ \hfill\cr
\hfill \quad \hbox{(3) ``Hot IGM''\hskip 0.7cm}\quad \hfill|
\hfill 0. \hfill|
\hfill 0. \hfill|
\hfill $0.000089$ \hfill|
\hfill $0.001105$ \hfill|
\hfill $0.034414$ \hfill|
\hfill $0.402218$ \hfill\cr
\hfill \quad \hbox{(4) ``Voids''\hskip 1.35cm}\quad \hfill|
\hfill 1. \hfill|
\hfill 1. \hfill|
\hfill $0.999854$ \hfill|
\hfill $0.998096$ \hfill|
\hfill $0.953032$ \hfill|
\hfill $0.585602$ \hfill
\endtable
\vfill\eject

\centerline {{\bf Table 2a.} Summary of
the volume weighted fractions in the $L=16h^{-1}$Mpc box model}
\medskip
\begintable
\hfill Redshift \hfill|
\hfill \quad 10\quad\hfill|
\hfill \quad 5\quad \hfill|
\hfill \quad 3\quad \hfill|
\hfill \quad 2\quad \hfill|
\hfill \quad 1 \hfill|
\hfill \quad 0 \quad \hfill\cr
\hfill \quad \hbox{(1) ``Virialized Gas''}\quad \hfill|
\hfill 0. \hfill|
\hfill $0.000001$ \hfill|
\hfill $0.000008$ \hfill|
\hfill $0.000083$ \hfill|
\hfill $0.000301$ \hfill|
\hfill $0.000713$ \hfill\cr
\hfill \quad \hbox{(2) ``Galaxies''\hskip 0.9cm}\quad \hfill|
\hfill 0. \hfill|
\hfill $0.000007$ \hfill|
\hfill $0.000044$ \hfill|
\hfill $0.000354$ \hfill|
\hfill $0.001347$ \hfill|
\hfill $0.001333$ \hfill\cr
\hfill \quad \hbox{(3) ``Hot IGM''\hskip 0.8cm}\quad \hfill|
\hfill 0. \hfill|
\hfill $0.000000$ \hfill|
\hfill $0.000001$ \hfill|
\hfill $0.000007$ \hfill|
\hfill $0.000179$ \hfill|
\hfill $0.008122$ \hfill\cr
\hfill \quad \hbox{(4) ``Voids''\hskip 1.35cm}\quad \hfill|
\hfill 1. \hfill|
\hfill $0.999992$ \hfill|
\hfill $0.999947$ \hfill|
\hfill $0.999556$ \hfill|
\hfill $0.998173$ \hfill|
\hfill $0.989832$ \hfill
\endtable
\bigskip
\bigskip
\bigskip
\bigskip
\bigskip

\centerline {{\bf Table 2b.} Summary of
the mass weighted fractions in the $L=16h^{-1}$Mpc box model}
\medskip
\begintable
\hfill Redshift\hfill|
\hfill \quad 10\quad\hfill|
\hfill \quad 5\quad \hfill|
\hfill \quad 3\quad \hfill|
\hfill \quad 2\quad \hfill|
\hfill \quad 1 \hfill|
\hfill \quad 0 \quad \hfill\cr
\hfill \quad \hbox{(1) ``Virialized Gas''}\quad \hfill|
\hfill 0. \hfill|
\hfill $0.000016$ \hfill|
\hfill $0.000101$ \hfill|
\hfill $0.001141$ \hfill|
\hfill $0.004796$ \hfill|
\hfill $0.014527$ \hfill\cr
\hfill \quad \hbox{(2) ``Galaxies''\hskip 0.9cm}\quad \hfill|
\hfill 0. \hfill|
\hfill $0.000099$ \hfill|
\hfill $0.000676$ \hfill|
\hfill $0.006354$ \hfill|
\hfill $0.030025$ \hfill|
\hfill $0.059315$ \hfill\cr
\hfill \quad \hbox{(3) ``Hot IGM''\hskip 0.8cm}\quad \hfill|
\hfill 0. \hfill|
\hfill $0.000000$ \hfill|
\hfill $0.000005$ \hfill|
\hfill $0.000118$ \hfill|
\hfill $0.010184$ \hfill|
\hfill $0.151940$ \hfill\cr
\hfill \quad \hbox{(4) ``Voids''\hskip 1.35cm}\quad \hfill|
\hfill 1. \hfill|
\hfill $0.999885$ \hfill|
\hfill $0.999217$ \hfill|
\hfill $0.992387$ \hfill|
\hfill $0.955071$ \hfill|
\hfill $0.774229$ \hfill
\endtable
\bigskip
\vfill\eject

\centerline {{\bf Table 3a.} Summary of
the volume weighted fractions in the $L=4h^{-1}$Mpc box model}
\medskip
\begintable
\hfill Redsift \hfill|
\hfill \quad 10\quad\hfill|
\hfill \quad 5\quad \hfill|
\hfill \quad 3\quad \hfill|
\hfill \quad 2\quad \hfill|
\hfill \quad 1 \hfill|
\hfill \quad 0 \quad \hfill\cr
\hfill \quad \hbox{(1) ``Virialized Gas''}\quad \hfill|
\hfill 0. \hfill|
\hfill $0.000025$ \hfill|
\hfill $0.000011$ \hfill|
\hfill $0.000010$ \hfill|
\hfill $0.000025$ \hfill|
\hfill $0.000018$ \hfill\cr
\hfill \quad \hbox{(2) ``Galaxies''\hskip 0.9cm}\quad \hfill|
\hfill 0. \hfill|
\hfill $0.000000$ \hfill|
\hfill $0.000006$ \hfill|
\hfill $0.000021$ \hfill|
\hfill $0.000056$ \hfill|
\hfill $0.000025$ \hfill\cr
\hfill \quad \hbox{(3) ``Hot IGM''\hskip 0.8cm}\quad \hfill|
\hfill 0. \hfill|
\hfill $0.000000$ \hfill|
\hfill $0.000000$ \hfill|
\hfill $0.000000$ \hfill|
\hfill $0.000001$ \hfill|
\hfill $0.001876$ \hfill\cr
\hfill \quad \hbox{(4) ``Voids''\hskip 1.35cm}\quad \hfill|
\hfill 1. \hfill|
\hfill $0.999995$ \hfill|
\hfill $0.999983$ \hfill|
\hfill $0.999969$ \hfill|
\hfill $0.999922$ \hfill|
\hfill $0.998081$ \hfill
\endtable
\bigskip
\bigskip
\bigskip
\bigskip
\bigskip

\centerline {{\bf Table 3b.} Summary of
the mass weighted fractions in the $L=4h^{-1}$Mpc box model}
\medskip
\begintable
\hfill Redshift\hfill|
\hfill \quad 10\quad\hfill|
\hfill \quad 5\quad \hfill|
\hfill \quad 3\quad \hfill|
\hfill \quad 2\quad \hfill|
\hfill \quad 1 \hfill|
\hfill \quad 0 \quad \hfill\cr
\hfill \quad \hbox{(1) ``Virialized Gas''}\quad \hfill|
\hfill 0. \hfill|
\hfill $0.000060$ \hfill|
\hfill $0.000163$ \hfill|
\hfill $0.000160$ \hfill|
\hfill $0.001175$ \hfill|
\hfill $0.003796$ \hfill\cr
\hfill \quad \hbox{(2) ``Galaxies''\hskip 0.9cm}\quad \hfill|
\hfill 0. \hfill|
\hfill $0.000000$\hfill|
\hfill $0.000506$ \hfill|
\hfill $0.003213$ \hfill|
\hfill $0.013870$ \hfill|
\hfill $0.034122$ \hfill\cr
\hfill \quad \hbox{(3) ``Hot IGM''\hskip 0.8cm}\quad \hfill|
\hfill 0. \hfill|
\hfill $0.000000$ \hfill|
\hfill $0.000000$ \hfill|
\hfill $0.000000$ \hfill|
\hfill $0.000007$ \hfill|
\hfill $0.012791$ \hfill\cr
\hfill \quad \hbox{(4) ``Voids''\hskip 1.35cm}\quad \hfill|
\hfill 1. \hfill|
\hfill $0.999940$ \hfill|
\hfill $0.999331$ \hfill|
\hfill $0.996627$ \hfill|
\hfill $0.984948$ \hfill|
\hfill $0.949294$ \hfill
\endtable
\bigskip
\vfill\eject

\centerline {{\bf Table 4.} Mass fraction and average mass of collapsed
objects in the four models}
\medskip
\begintable
\hfill \quad L ($h^{-1}$Mpc)\quad \hfill|
\hfill \quad 64\quad \hfill|
\hfill \quad 16\quad \hfill|
\hfill \quad 4\quad \hfill\cr
\hfill \quad $<m>$\quad \hfill|
\hfill \quad $5.9\times 10^{10} M_\odot$\quad \hfill|
\hfill \quad $2.5\times 10^9 M_\odot$\quad \hfill|
\hfill \quad $1.3\times 10^9 M_\odot$\quad \hfill\cr
\hfill f(collasped) \hfill|
\hfill $0.012$ \hfill|
\hfill $0.030$ \hfill|
\hfill $0.014$ \hfill
\endtable
\bigskip
\vfill\eject

\centerline {{\bf Table 5.}
The mean Zeldovich-Sunyaev $y$ parameter as a function of redshift}
\medskip
\begintable
\hfill Redshift \hfill|
\hfill 5 \hfill|
\hfill 3 \hfill|
\hfill 2 \hfill|
\hfill 1 \hfill|
\hfill 0.5 \hfill|
\hfill 0 \hfill\cr
\hfill $\bar y$ \hfill|
\hfill $5.0\times 10^{-12}$ \hfill|
\hfill $1.5\times 10^{-10}$ \hfill|
\hfill $6.0\times 10^{-10}$ \hfill|
\hfill $1.1\times 10^{-8}$ \hfill|
\hfill $2.7\times 10^{-8}$ \hfill|
\hfill $2.7\times 10^{-7}$ \hfill
\endtable
\medskip
\vfill\eject
\end